\documentclass[preprint,showpacs,superscriptaddress,longbibliography]{revtex4-2}
\usepackage{amsmath,amssymb,bm}
\usepackage{mathrsfs}
\usepackage{textcomp}
\usepackage{commath}
\usepackage{mathtools}

\usepackage{hyperref}
\hypersetup{
    colorlinks=true,
    linkcolor=blue,
}

\usepackage{natbib}
\usepackage{placeins}
\usepackage[final]{graphicx} 
\usepackage{rotating}
\usepackage{epstopdf}
\usepackage{float}
\usepackage{amsmath}
\usepackage[lofdepth=1,lotdepth]{subfig}
\usepackage[usenames]{color} 

\usepackage{cancel} 

\usepackage[normalem]{ulem}
\renewcommand{\sout}[1]{} 

\newcommand{\Ca}{\mathrm{Ca}}
\newcommand{\kb}{\hat{K}_B}

\graphicspath{{figures/}}  

\makeatletter

\newcommand{\numtoRoman}[1]{\expandafter\@slowromancap\romannumeral #1@}
\makeatother

\begin{document}

\title{Wrinkling and multiplicity in the dynamics of deformable sheets in uniaxial extensional flow}
\author{Yijiang Yu}
\affiliation{
Department of Chemical and Biological Engineering\\
University of Wisconsin-Madison, Madison, WI 53706-1691
}
\author{Michael D. Graham}\email{Corresponding author. E-mail: mdgraham@wisc.edu}
\affiliation{
Department of Chemical and Biological Engineering\\
University of Wisconsin-Madison, Madison, WI 53706-1691
}
\date{\today}

\begin{abstract}
The processing of thin-structured materials in a fluidic environment, from nearly inextensible but flexible graphene sheets to highly extensible polymer films, arises in many applications. So far, little is known about the dynamics of such thin sheets freely suspended in fluid. In this work, we study the dynamics of freely suspended soft sheets in uniaxial extensional flow.
Elastic sheets are modeled with a continuum model that accounts for in-plane deformation and out-of-plane bending, and the fluid motion is computed using the method of regularized Stokeslets. 
We explore two types of sheets: ``stiff" sheets that strongly resist bending deformations and always stay flat, and ``flexible" sheets with both in-plane and out-of plane deformability that can wrinkle. 
For stiff sheets, we observe a coil-stretch-like transition, similar to what has been observed for long-chain linear polymers under extension as well as elastic sheets under planar extension: in a certain range of capillary number (flow strength relative to in-plane deformability), the sheets exhibit either a compact or a highly stretched conformation, depending on deformation history.
For flexible sheets, sheets with sufficiently small bending stiffness wrinkle to form various conformations. Here, the compact-stretched bistability still occurs, but is strongly modified by the wrinkling instability: a highly-stretched planar state can become unstable and wrinkle, after which it may dramatically shrink in length due to hydrodynamic screening associated with wrinkling. Therefore, wrinkling renders a shift in the bistability regime. In addition, we can predict and understand the nonlinear long-term dynamics for some parameter regimes with linear stability analysis of the flat steady states.
\end{abstract}
\maketitle

\section{INTRODUCTION} \label{sec:introduction}
   
Two-dimensional materials, such as graphene, boron nitride and thin polymeric films, have received great attention in recent years.
Thin graphene flakes and graphene composites, due to their superior mechanical and electronic properties, are widely used in the energy storage field \cite{Fei2016,alamer2020, zamani2021ultralight, zhang2011high}. Electronic nanomaterials, such as $\mathrm{MoS_2}$,  $\mathrm{GeSe}$ and graphene, are used in nanoscale thin films for semiconductor nanomembranes in high-performance electronic devices \cite{Rogers2011,Yoon2010}. Thin polymer sheets and films are important in the development of cell-based biohybrid machines \cite{Liu2016}. 

These examples all involve stages of synthesis or processing where thin sheet-like structures are freely suspended in a fluidic environment, potentially showing complicated dynamics under external flow fields. 
For instance, Fei and coworkers improved the performance of carbon nanofibers in battery anodes by adding graphene oxide (GO) nanosheets to the fibers, which were produced by water-based electrospinning \cite{Fei2016}. Once the mixture of nanosheets and polymer solutions was injected into the air, the uniaxial extensional flow near the needle tip stretched the nanosheet and helped it to wrap around the fiber surface.
Park $\textit{et al.}$ investigated behavior and orientation of layered silicate nanocomposites in uniaxial extensional flow, finding that exfoliated nanosheets tended to align with the extensional direction \cite{park2006rheological}.
Recently, Ng $\textit{et al.}$ performed capillary breakup extensional rheometry (CaBER) with dilute GO solutions, demonstrating that a very small amount of GO nanosheets can influence extensional rheology \cite{ng2020go}.
Yoon $\textit{et al.}$ synthesized GaAs nanomembranes in suspension with a combination of dispersion and sonication that generated complicated flow fields \cite{Yoon2010}.
Liu $\textit{et al.}$ reviewed the fabrication of thin and ultra-thin polymer films, some of which involved an exfoliation process to dissolve the substrate where the films were suspended in solution \cite{Liu2016}. 

The understanding of these nontrivial scenarios requires a comprehensive knowledge of the dynamics of suspended deformable sheet-like structures in flow, an area which has barely been explored.
The above examples involve uniaxial extensional flow, which is therefore the focus of the present work. To our knowledge, this is the first study of flexible sheets in uniaxial extension.

Though little is known about deformable sheet dynamics in extensional flow, other soft objects such as  flexible polymers have been well-studied.
De Gennes suggested the possibility of a discontinuous, hysteretic coil-stretch transition for sufficiently long chain polymers under extensional flow, because of the conformation-dependence of the hydrodynamic interactions between chain segments \cite{de1974}.
Fuller and Leal modeled this situation in planar extensional flow with a bead-spring dumbbell model with conformation-dependent drag on the beads, observing the discontinuous transition and conformational bistability and hysteresis hypothesized by De Gennes \cite{fuller1981}.
Bistability was experimentally observed by Schroeder and coworkers with dilute solutions of fluorescently labeled E. coli DNA under planar extension \cite{schroeder2003}. 
They also performed Brownian dynamics simulations of a bead-spring polymer model with hydrodynamic interactions, finding conformational hysteresis consistent with what they observed in experiments.
Recent computational work has also suggested the transition not only happens in a dilute polymer solution but also for polyethylene melts under planar extension \cite{nafar2018communication}.

Coil-stretch-like transitions have been shown to exist beyond linear polymers. Kantsler $\textit{et al.}$ reported experimental observation of a tubular-dumbbell transition for vesicles in planar elongational flow when a critical strain rate is reached \cite{kantsler2008}. They did not observe bistability.
Zhao $\textit{et al.}$ found a similar transition computationally for a vesicle in uniaxial extensional flow \cite{zhao2013shape}: when reaching a critical flow strength, the vesicle became unstable and deformed into an extended asymmetric dumbbell shape with two unequally sized ends. The simulation results are consistent with experimental observations from Kantsler $\textit{et al.}$ in planar extension and vesicles did not show bistability. 
Soh and Doyle studied the deformation response for 2D kinetoplast DNA sheets, which are networks of linked ring structures that appear in mitochondria, in planar elongational flow \cite{soh2020deformation}. The DNA sheets underwent a relatively smooth transition from an undeformed state to a deformed stretched state, with no evidence of a hysteretic coil-stretch transition. 
A recent computational study from Yu and Graham demonstrated that a sufficiently deformable elastic sheet in planar or biaxial extension can display a compact-stretched transition with hysteresis that is closely analogous to the coil-stretch transition in long polymer chains in solution \cite{yu2021coil}.
As in the polymer case, the bistability was shown to arise from the hydrodynamic interaction between different parts of the sheet surface, and vanished if the hydrodynamic interactions were turned off. 
Due to the nature of biaxial and planar extension, the long-time conformations of the sheets were always flat. However, as we shall see, a flexible sheet in uniaxial extension experiences radial compressive stress and may wrinkle, influencing the shape evolution and bistability.

While the above discussion focused on extensional flows, a number of recent studies consider the dynamics of inextensible sheets in shear flow in order to understand synthesis processes of sheets such as graphene, which usually involve a liquid-phase exfoliation in a shear-dominated flow where complicated folding dynamics may occur.
Xu and Green studied a Brownian bead-spring model for a square sheet in simple shear flow \cite{xu2014}, where they characterized a cyclic crumple-stretch-crumple motion for a flexible sheet. In addition, they also applied the same model to investigate the dynamics in biaxial extension \cite{xu2015}, where sheets with large bending stiffness stayed flat while sheets with small stiffness crumpled as a result of Brownian fluctuations.
Silmore and coworkers studied semiflexible non-Brownian sheets in shear  \cite{silmore2021buckling}. They found that, depending on initial conditions and bending stiffness, the sheets can either undergo a continuous chaotic tumbling or follow a Jeffery-like orbit then reach a flat state in the flow-vorticity plane. They also predicted some transient buckling dynamics with a simple 1D elasticity model. 
Dutta and Graham studied piecewise rigid creased sheets in shear flow \cite{dutta2017}. Sheets either approached a steady, periodic or quasiperiodic state depending on initial conditions. 
Botto and coworkers examined liquid-phase exfoliation of multiple graphene sheets via molecular dynamic simulation and characterized the parameter regime where exfoliation can occur \cite{gravelle2020liquid}. They also applied molecular dynamic simulation to study thin nanoplatelets in shear \cite{kamal2020hydrodynamic}. They found the effect of hydrodynamic slip influenced the dynamics, allowing the platelet to attain a stable steady orientation with a fixed angle with respect to flow.

In the present work, we investigate the dynamics of soft, non-Brownian elastic sheets in uniaxial extensional flow. We examine cases that range from relatively stiff sheets like graphene to highly deformable polymeric sheets. We are interested in the wrinkling instability and its influence on compact-stretched bistability -- i.e.~the analogue of the coil-stretch transition in linear polymers. The rest of the paper is presented as follows: Section \ref{sec:methods} presents the model and numerical methods. Section \ref{sec:results} presents the results in three parts: dynamics of stiff sheets (sheets with large stiffness that prevents out-of-plane deformation); wrinkling transitions at low extension rates where sheets are not highly stretched; dynamics of stretched sheets and the interaction between wrinkling and the compact-stretched transition. We conclude in Section \ref{sec:conclusion}.

\section{MODEL DEVELOPMENT} \label{sec:methods}
     
\subsection{Model setup}

We consider an elastic thin sheet that is freely suspended in an unbounded uniaxial extensional flow ($\mathbf{v}_\infty = \dot\varepsilon\left[-x/2, -y/2, z \right]^T$, with $\dot\varepsilon$ as the flow strain rate). The fluid is Newtonian with viscosity $\eta$ and density $\rho$. We model the sheet as a continuum with a no-slip surface, so material points on the surface move with the same velocity as the local fluid.
The no-slip boundary condition also implies that that the sheet is impermeable -- if the sheet moves with the local fluid velocity, then no fluid is passing through the sheet.
The sheet surface is discretized into triangular elements with a node at each corner. The detailed numerical methods will be introduced in the following section. In the present study, we focus on a sheet whose equilibrium shape is a disc with radius $a$ and thickness $h$ such that $h \ll a$. Thus, the sheet has a traction-free edge, along which fluid exerts no forces. 

The total strain energy of the elastic sheet can be written as the sum of in-plane  strain energy $E_s$ and out-of plane bending energy $E_b$:
\begin{equation}
    E = E_s + E_b.
\end{equation}
Here, the in-plane strain energy is evaluated by integrating the strain energy density $W$ over the sheet surface. We choose a nonlinear Yeoh model (YH), as it is often used to model rubber-like polymer structures \cite{yeoh1993}, with the form:

\begin{equation}
    W_\mathrm{YH} = \frac{G}{2}\left( I_1 -3\right) + cG\left( I_1 -3\right)^3.
\end{equation}
The two-dimensional shear modulus $G$ scales with the equilibrium thickness $h$. The term $I_1$ is the first invariant of the right Cauchy-Green deformation tensor, and depends on the local principal stretch ratios $\lambda_i$ along the tangential direction of the sheet surface:

\begin{equation}
	I_{1}=\lambda_1^2+\lambda_2^2+\frac{1}{\lambda_1^2\lambda_2^2}.
\end{equation}
The last term denotes $\lambda_3$, as the stretch ratio along the thickness direction, is obtained by incompressibility of the material ($\lambda_1\lambda_2\lambda_3 = 1$). The parameter $c$ weights the cubic term in the energy density, and can be physically related to the properties of the material. For example, for a polymeric sheet, $c$ is inversely related to the distance between cross-linked points inside the sheet, thus small $c$ indicates higher extensibility. Consequently, large $c$ makes the material strain-hardening, due to the penalty from the cubic energy term under large strain. If $c = 0$, the model recovers the strain-softening neo-Hookean (NH) model.

For the bending energy $E_b$, we apply a simple bending model that sums the energy due to dihedral angles $\theta_{\alpha\beta}$ between neighboring elements \cite{Fedosov2010,fedosov2010multiscale}:
\begin{equation}
    E_b = \sum_{\mathrm{adj}\ \alpha,\beta}k_b \left(1-\cos(\theta_{\alpha\beta} - \theta_0)\right)
\end{equation}
where $k_b$ is the bending constant calculated from bending stiffness $K_B$ ($k_b = \frac{3\sqrt{3}}{2}K_B$). The angle $\theta_{0}$ is zero as we assume a flat equilibrium state. 

In addition, we also add a truncated Lennard-Jones(LJ) repulsive potential to prevent the sheet from self-intersecting. 
Numerically, the potential acts between all the nodes, except for the nodes that are located within 3-ring neighbors of the targeted node at the equilibrium state. The potential has the form:
\begin{equation}
	E_{LJ}= 
	\begin{cases}
		4 \varphi_0\left[\left(\frac{\sigma}{r}\right)^{12}-\left(\frac{\sigma}{r}\right)^{6}\right],	&\text{if } r<\sigma\\
		0,	&\text{otherwise}
	\end{cases}
\end{equation}
where $r$ is the distance between nonadjacent nodes at equilibrium state, $\sigma$ is the range of the potential, and $\varphi_0$ measures is the strength of the potential. Here, we choose $\sigma = 0.06a$ and $\varphi_0 = 4\times10^{-6}$ such that the potential has no influence on the dynamics for sheets without intersection. We find that the only case where the repulsive force is active is when the sheet edge would otherwise cut through the surface, which only occurs for sheets with very small bending stiffness.

We introduce two nondimensional parameters to describe the aforementioned mechanical properties. The capillary number $\Ca = \eta\dot\varepsilon a/G$ compares the viscous stresses exerted by the fluid on the sheet to the in-plane elastic response of the sheet. Larger $\Ca$ indicates that the sheet is more deformable. The relative size of the bending and in-plane stresses is measured by $\kb = K_B/a^2G$: the parameter comparing the bending with flow is $\Ca/\kb=\eta\dot{\epsilon}a^3/K_B$: smaller $\kb$ indicates that the sheet is more flexible. 
{In this study, we explore two types of sheets: ``stiff" sheets ($\kb = \infty$) with in-plane deformability, which strongly resist bending deformations and always stay flat, and ``flexible" sheets with both in-plane and out-of plane deformability, which can wrinkle. We consider the stiff sheet case primarily as an idealized base case for comparison with the more realistic flexible case.}

{Additionally, we do not include thermal fluctuations, as we estimate them to be negligible under the conditions studied here. For example, for a polyethylene glycol (PEG) hydrogel \cite{gaharwar2011highly} sheet with radius of 100 $\mathrm{\mu m}$ and thickness of 100 nm suspended in water, its estimated shear strain energy ($\sim 10^9 k_BT$) and bending energy ($\sim 3\times10^5 k_BT$) are much larger than thermal energy.}

\subsection{Numerical methods}
The numerical method for the elasticity problem is adapted from Charrier $\textit{et al.}$ \cite{Charrier1989,Pappu:2008in}. We simulate the sheet by keeping track of nodes that move as material points on the sheet surface. 
From the above two energies, we obtain the summed nodal elastic force ($\mathbf{F}_{e,i} = \mathbf{F}_{s,i}+\mathbf{F}_{b,i}$) exerted on each node from the first variation of the total energy with respect to the nodal displacements.
Each discretized element of the sheet is assumed to have homogenous deformation, so the element edges always remain linear. The deformed element is compared to its equilibrium shape under a local coordinate transformation via a rigid body rotation, and the displacement for any point inside the element is obtained by linear interpolation from the nodal positions. 
The detailed implementation can be found in references \cite{Kumar:2012ev,Fedosov2010,fedosov2010multiscale}. 
The total nodal force $\mathbf{F}_{e,i}$ is evaluated by summing the elastic force $(\mathbf{F}_{e,i})_j$ due to deformation of each surrounding element $j$ shared by the node $i$: $\mathbf{F}_{e,i} = \sum_j (\mathbf{F}_{e,i})_j$, where the sum is over all elements meeting at the node.
For the results shown, we discretize the disc with 1600 elements and 841 nodes. We have verified that changes in mesh resolution lead to only small quantitative changes in the results and no qualitative changes.

In the present work, we consider a small, thin sheet and assume that the particle Reynolds number $\mathrm{Re}=\rho\dot\epsilon a^2/\eta$ is negligible. Under these assumptions, inertia of the sheet and fluid are negligible, so the forces on each point on the sheet sum to zero and the fluid is governed by the Stokes equation. For each node on the sheet surface, the elastic force $\mathbf{F}_{e,i}$ exerted by the sheet on the fluid is balanced with hydrodynamic force $\mathbf{F}_{h,i}$ from the fluid:
\begin{equation}
    \mathbf{F}_{e,i} + \mathbf{F}_{h,i} =\mathbf{0}.
\end{equation} 

To account for the fact that the forces are not completely localized to the nodal positions, we use the method of regularized Stokeslets to solve for the fluid motion \cite{Cortez2005}: the force $\mathbf{F}_i = -\mathbf{F}_{h,i}$ exerted by node $i$ on the fluid corresponds to a regularized force density $\mathbf{f}^\kappa_i = \mathbf{F}_i \delta_\kappa(\mathbf{x}-\mathbf{X}_i)$, where $\mathbf{X}_i$ is the position of node $i$ and $\delta_\kappa(\mathbf{x})$ is a regularized delta function with regularization parameter $\kappa$. Thus, the governing equations are the Stokes equation with regularized nodal forces and the continuity equation:
\begin{equation}
    \begin{array}{c}
        {-\nabla p+\eta \nabla^{2} \mathbf{v}+\sum_i\mathbf{F}_i \delta_\kappa(\mathbf{x}-\mathbf{X}_i)=\mathbf{0}} \\ 
        {\nabla \cdot \mathbf{v}=0}.
    \end{array}
\end{equation}
The velocity field generated due to a regularized point force $\mathbf{f}= \mathbf{F}\delta_\kappa(\mathbf{x}-\mathbf{X}_i)$ can be represented using a regularized Stokeslet $\mathbf{G}_\kappa$:
\begin{equation}
    \mathbf{v}_\kappa(\mathbf{x}) = \mathbf{G}_\kappa(\mathbf{x}-\mathbf{X}_i)\cdot \mathbf{F}.
\end{equation}
As $1/\kappa\rightarrow 0$, $\mathbf{G}_\kappa$ reduces to the usual Stokeslet operator $\mathbf{G}(\mathbf{x}-\mathbf{X}_i)=1/8\pi\eta r(\mathbf{I}+(\mathbf{x}-\mathbf{X}_i)(\mathbf{x}-\mathbf{X}_i)/r^2)$, where $r=||\mathbf{x}-\mathbf{X}_i||$.
There are many ways to regularize a delta function $\delta_\kappa(\mathbf{x})$; we choose a regularization function for which the difference between $\mathbf{G}$ and $\mathbf{G}_\kappa$ decays exponentially as $\kappa r\rightarrow\infty$ \cite{hernandez2007,Graham2018, nguyen2014reduction}:

\begin{equation}
    \delta_\kappa(\mathbf{x}-\mathbf{X}_i) = \frac{\kappa^3}{\sqrt{\pi}^3}\exp(-\kappa^2r^2)\left[\frac52-\kappa^2r^2\right].
\end{equation}
With this choice, 
\begin{equation}
    \mathbf{G}_{\kappa}(\mathbf{x}-\mathbf{X}_i)=\frac{\operatorname{erf}(\kappa r)}{8 \pi \eta r}\left(\mathbf{I}+\frac{(\mathbf{x}-\mathbf{X}_i)(\mathbf{x}-\mathbf{X}_i)}{r^{2}}\right)+\frac{\kappa e^{-\kappa^{2} r^{2}}}{4 \pi^{3 / 2} \eta}\left(\mathbf{I}-\frac{(\mathbf{x}-\mathbf{X}_i)(\mathbf{x}-\mathbf{X}_i)}{r^{2}}\right).
\end{equation}
In the simulations, $\kappa$ must be chosen to scale with the minimum node-to-node distance $l_{\mathrm{min}}$. We take $\kappa l_{\mathrm{min}} = 2.1842$, which is obtained from a validation case of a disc in biaxial extensional flow discussed in \cite{yu2021coil}. We represent the total velocity at a point $\mathbf{x}$ as 
\begin{equation}
    \mathbf{v}(\mathbf{x}) =\mathbf{v}_\infty(\mathbf{x}) +\mathbf{v}_p(\mathbf{x})= \mathbf{v}_\infty(\mathbf{x}) +\sum_i\mathbf{G}_\kappa(\mathbf{x}-\mathbf{X}_i)\cdot \mathbf{F}_i.
    \label{eq:v}
\end{equation}

At each time step, after updating the force balance, we update the nodal positions by applying the condition that they move with the local fluid velocity in Eq.~\ref{eq:v}, thereby satisfying the no-slip and no-penetration conditions:
\begin{equation}
	\frac{d\mathbf{X}_i}{dt}=\mathbf{v}(\mathbf{X}_i).
\end{equation}

This equation is solved with a fourth-order Runge-Kutta method.
For numerical stability, the time step applied follows $\Delta t = 0.1\Ca l_{\mathrm{min}}$. If $\Ca$ is large, we apply an upper limit of $5 \times 10^{-4}$ strain unit per time step.

\begin{figure}[t!]
    \centering
    \captionsetup{justification=raggedright}
    \includegraphics[width=0.7\textwidth]{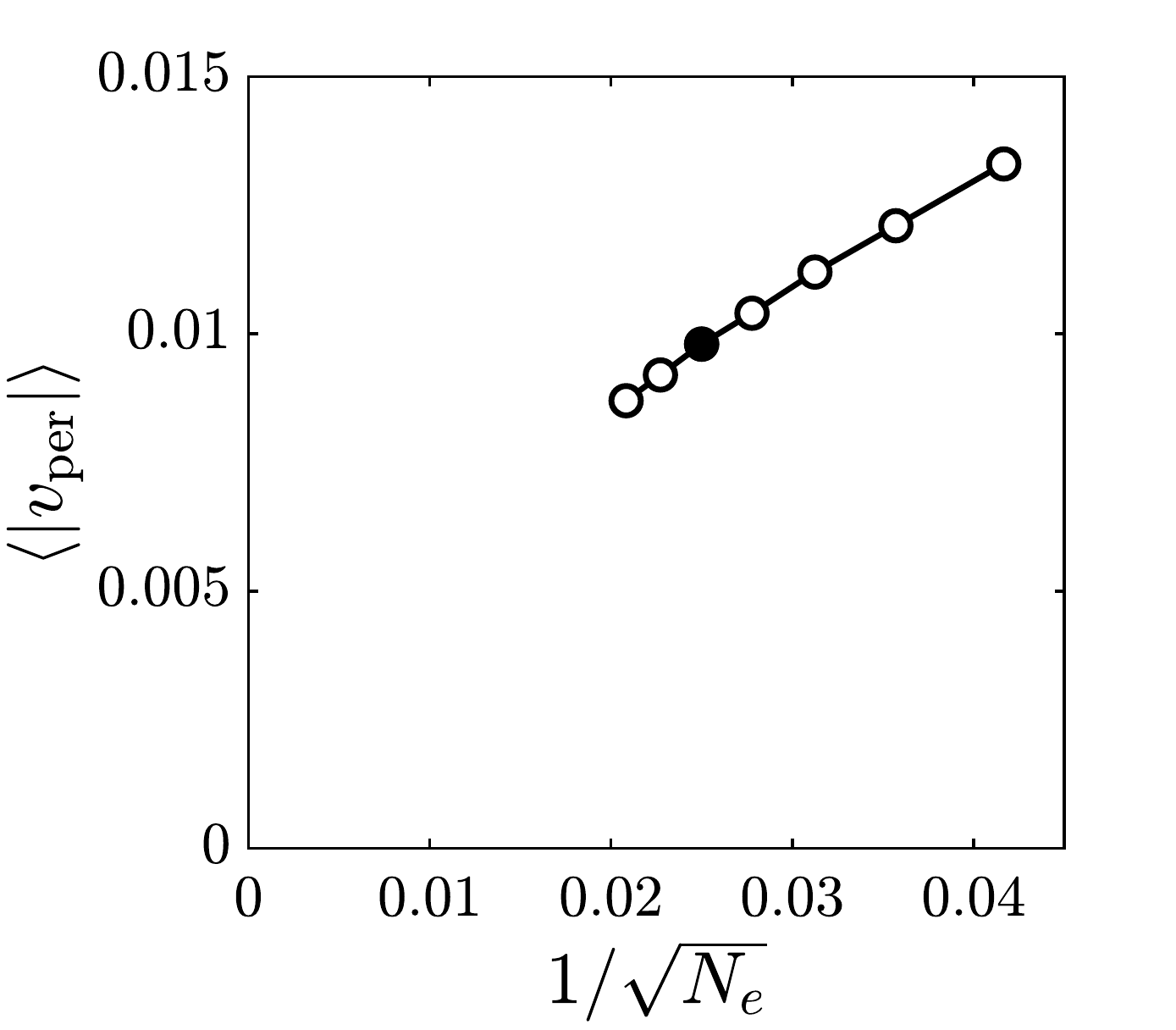} 
    \caption{Average magnitude of permeation velocity, $\langle|v_\mathrm{per}|\rangle$, at element centers for a sheet (the taco shape in Fig.~\ref{fig:resc_flex_evo}(a)) vs.~mesh size 1/$\sqrt{N}_e$. The filled symbol indicates the mesh of 1600 triangular elements used for the main set of results.}
    \label{fig:model_vn}
\end{figure}

Although we satisfy the no-slip and no-penetration conditions to high accuracy on the nodal points $\mathbf{X}_i$, with a finite discretization these conditions cannot be exactly satisfied at every point on the sheet. 
In particular, with a finite-resolution mesh, the sheet may be slightly permeable, so  we address here the effect of resolution on the velocity normal to the sheet surface relative to the velocity of the sheet itself. At the center of each triangular element, we can define two velocities: $\mathbf{v}_\mathrm{int}$ is the velocity estimated by linear interpolation from the velocities on the nodes, and $\mathbf{v}_\mathrm{num}$ is the velocity obtained from the regularized Stokeslet expression, Eq.~\ref{eq:v}, evaluated at that point. At that point the permeation velocity $v_\mathrm{per}$, i.e.~the error in the no-penetration condition, is given by 
\begin{equation}
v_\mathrm{per}=\mathbf{n}\cdot(\mathbf{v}_\mathrm{int}-\mathbf{v}_\mathrm{num}),
\end{equation}
where $\mathbf{n}$ is the unit normal of the element. To evaluate $v_\mathrm{per}$, we use as an example the steady state sheet conformation shown below in Fig.~\ref{fig:resc_flex_evo}(a), and note that the linear size of each element in a mesh with $N_e$ elements scales as $1/\sqrt{N_e}$. Figure \ref{fig:model_vn} shows the averaged magnitude of perturbation velocity $\langle|v_\mathrm{per}|\rangle$ vs.~$1/\sqrt{N_e}$ for this case. The result indicates that the error is small and tends roughly linearly toward zero as the mesh gets finer. The choice of mesh size (filled symbol on the plot) used for the main results gives about 1\% error for points at the element centers, and as noted above, we perfectly satisfy no-slip and no-penetration conditions on the nodal positions $\mathbf{X}_i$.

Finally, in all the results reported here, we examine the dynamics of the sheet based on a local Lagrangian frame centered on the sheet. Because of the saddle point nature of the streamlines near the stagnation point in uniaxial extensional flow, any small perturbation or symmetry-breaking of the sheet position or orientation will cause it to move away from the stagnation point. Nevertheless, as in any linear flow, the sheet sees the same velocity gradient everywhere, so the shape dynamics, which are our focus here, are not influenced by the Eulerian position of the sheet.

\section{RESULTS AND DISCUSSION} \label{sec:results} 
    \subsection{Dynamics of stiff sheets ($\kb = \infty$)} \label{sec:results_stiff}

We begin the discussion of dynamics in uniaxial extensional flow by introducing the dynamics of a stiff sheet that has extremely large bending stiffness ($\kb = \infty$) and always stays flat. The dynamics of a stiff sheet serve as a base case to understand the more complicated flexible sheets discussed in later sections. For simplicity, we impose the infinite stiffness criterion by directly preventing any out-of-plane deflection in the numerical simulation. Our primary quantity for characterizing the conformation of a sheet will be its half-length $l$ in the $z$-direction, which at rest reduces to its radius $a$.

One important observation of a sheet in uniaxial flow is that the only stable orientation of the sheet aligns with the flow direction. This observation agrees with Park $\textit{et al.}$ \cite{park2006rheological} for orientation of nanocomposites in uniaxial extension. Therefore, we only focus on the initial condition where the sheet aligns with flow direction: the sheet always lies in the $x-z$ plane.

\begin{figure*}[!t]
    \centering
    \captionsetup{justification=raggedright}
    \includegraphics[width=\textwidth]{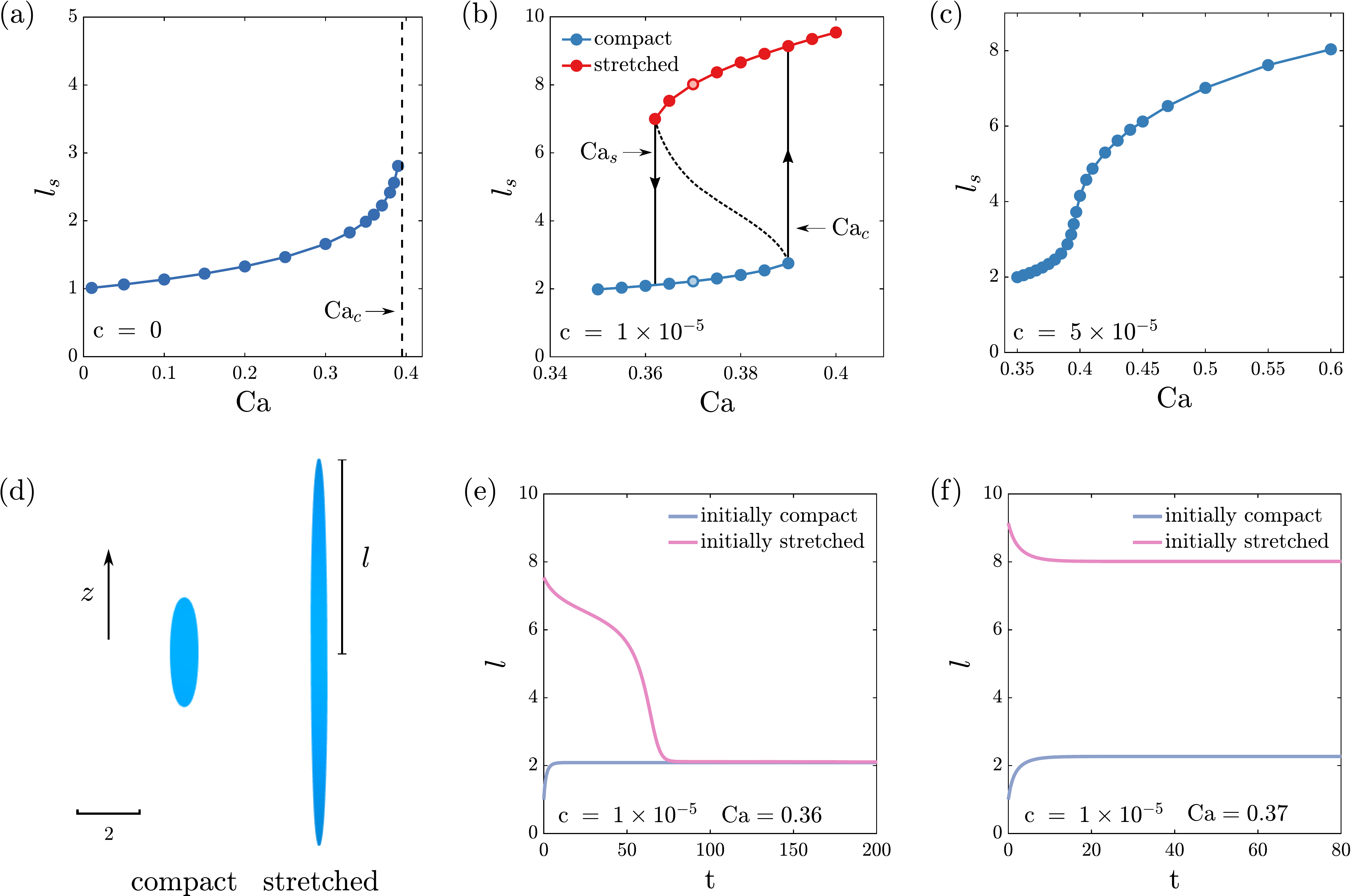}
    \caption{(a) {Steady state half-length} $l_s$ vs. $\Ca$ for a stiff disc with $c = 0$ (neo-Hookean) in uniaxial extensional flow, indicating a singularity in $l_s$. (b) $l_s$ vs. $\Ca$ for a stiff disc with $c = 1\times 10^{-5}$ in uniaxial extensional flow. Blue symbols represent a compact final conformation and red symbols represent a stretched final conformation. $\Ca_c$ and $\Ca_s$ are the two critical $\Ca$ marked beside the arrows. (c) $l_s$ vs. $\Ca$ for a stiff disc with $c = 5\times 10^{-5}$ in uniaxial extensional flow, with a smooth transition observed. (d) Examples of bistable sheets for $\Ca = 0.37$ with $c = 1\times 10^{-5}$, corresponding to light symbols in Fig.~\ref{fig:resu_stiff_csh}(b). The figure shows the front view of compact and stretched states with half-length labeled $l_s$. (e) Examples of transient length evolution for stiff sheets with $\Ca = 0.36$ and $c = 1\times 10^{-5}$, below the bistability regime. (f) Examples of transient length evolution for stiff sheets with $\Ca = 0.37$ and $c = 1\times 10^{-5}$ , in the bistable regime.}
    \label{fig:resu_stiff_csh}
\end{figure*}

Figure \ref{fig:resu_stiff_csh}(a) - (c) presents bifurcation diagrams for the steady state value $l_s$ vs.~$\Ca$ with different choices of parameter $c$, with the definition of $l$ shown in Fig.~\ref{fig:resu_stiff_csh}(d). 
For a simple neo-Hookean model ($c = 0$), shown in Fig.~\ref{fig:resu_stiff_csh}(a), the conformation stretches without bound once a critical capillary number $\Ca_c = 0.39$ is exceeded, yielding a singularity. (This is analogous to the unbounded stretching of a bead-spring dumbbell of a polymer in solution when the spring is Hookean \cite{Graham2018}.) For a nonlinear Yeoh model with small but finite $c$ (Fig.~\ref{fig:resu_stiff_csh}(b)), beyond $\Ca_c$, the compact state loses existence, and the sheets evolve to a stretched state -- the singularity in the Hookean case at $\Ca_c$ is replaced by a saddle-node bifurcation at which the compact branch loses existence. Quasistatically decreasing $\Ca$ shows that the stretched steady state branch persists down to another critical (saddle-node bifurcation) value $\Ca_s$, below which all initial conditions relax back to the compact state. That is, when $\Ca_s<\Ca<\Ca_c$, this case displays bistability between a weakly stretched compact branch and a fully stretched branch. 
Figure \ref{fig:resu_stiff_csh}(d) shows examples of compact and stretched conformations for $\Ca = 0.37$ from Fig.~\ref{fig:resu_stiff_csh}(b). This bifurcation behavior is qualitatively similar to what has been previously predicted for sheets in planar and biaxial extension \cite{yu2021coil}.

{We compare two examples from Fig.~\ref{fig:resu_stiff_csh}(b) to show the difference in dynamics for sheets with and without bistability}: Figure \ref{fig:resu_stiff_csh}(e) shows the transient length evolution of sheets for a value of $\Ca$ below the bistable regime, as either a compact or stretched initial condition both result in a compact final conformation as the only stable steady state; Figure \ref{fig:resu_stiff_csh}(f) indicates that if $\Ca$ is inside the bistable regime, the sheet either becomes compact or remains stretched based on its initial condition, with both states being stable.

The bistability regime depends on the value of $c$. Smaller $c$ indicates larger extensibility, with $c = 0$ the special case of neo-Hookean elasticity. The bistability arises only when $c$ is small. 
Figure \ref{fig:resu_stiff_csh}(c) shows an example where $c$ is sufficiently large that $l_s$ increases smoothly with increasing $\Ca$. (As $c$ increases, $\Ca_s$ approach one another, eventually merging and disappearing.)  Recalling that $c$ can be viewed as inversely related to the extensibility of the polymer subchains in the sheet, we suspect that the reason that bistability was not observed in the DNA networks of \cite{soh2020deformation} is the relatively low molecular weight of the DNA that they used, leading effectively to a value of $c$ above the bistability regime. 
In addition, we note that changing $c$ has negligible influence on the stretching dynamics of the compact branch, because the strain is relatively small for compact conformations.  
The current study focuses on the compact-stretched transition, so below we only discuss the influence of deformation on the transition with choices of $c$ that show bistability ($c = 1\times 10^{-5}$). 

    \subsection{Dynamics of flexible sheets in the compact (low $\Ca$) regime} 
      \label{sec:results_compact}

In this section, we probe into the compact branch shown in Fig.~\ref{fig:resu_stiff_csh}(b). Here, we remove the infinite stiffness constraint to make the sheet flexible, so the dynamics depend on bending stiffness $\kb$.
When the sheet is compact, the dynamics are not sensitive to $c$, as the penalty from cubic energy term is small. Therefore, we present the following results with a simple neo-Hookean model ($c = 0$), where a singularity appears beyond $\Ca_c = 0.39$.

\begin{figure}[t!]
    \centering
    \captionsetup{justification=raggedright}
    \includegraphics[width=0.7\textwidth]{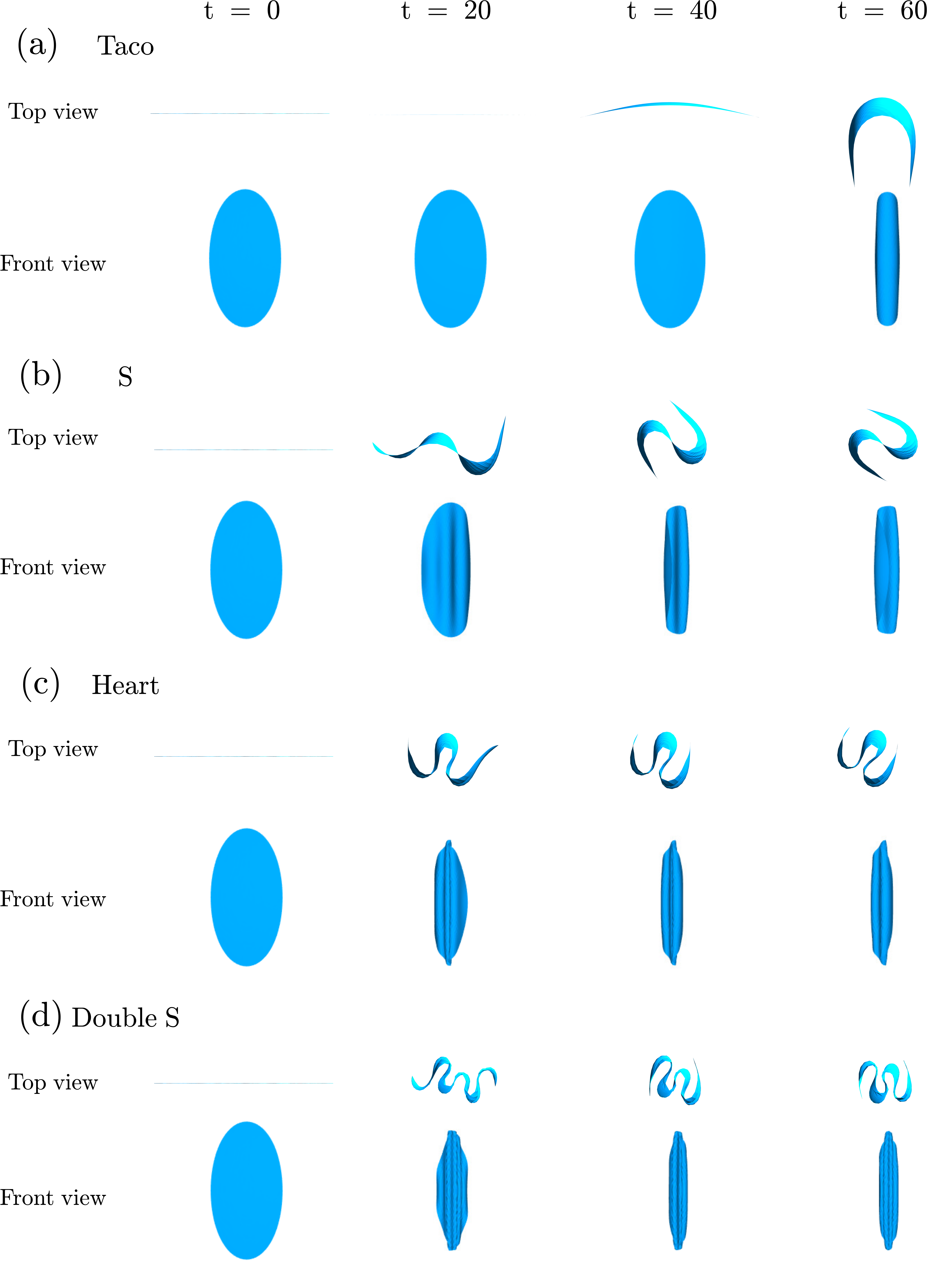} 
    \caption{Top view ($x-y$ plane) and front view ($x-z$ plane) of shape evolutions for a deformable sheet with $\Ca = 0.25$ and decreasing $\kb$: (a) Taco ($\kb = 5 \times 10^{-3}$), (b) S shape ($\kb = 1 \times 10^{-3}$), (c) Heart ($\kb = 3 \times 10^{-4}$), (d) Double S ($\kb = 1 \times 10^{-4}$). See Supplemental Material at [URL will be inserted by publisher: Appendix \ref{sec:appendix} Movie 1] for animated movies.}
    \label{fig:resc_flex_evo}
\end{figure}

The dynamics of flexible sheets in uniaxial flow are complicated due to flow-induced out-of-plane deformation and are highly dependent on $\Ca$ and $\kb$. We perform  numerical simulations by applying a small and random out-of-plane perturbation to the flat steady states shown in Fig.~\ref{fig:resu_stiff_csh}(b). We found that, depending on $\kb$, the sheet either resists perturbation and stays flat, or develops into different wrinkled conformations. 
Figure \ref{fig:resc_flex_evo} shows a transition of final conformation, for a sheet with fixed $\Ca$ and decreasing $\kb$, from a flat stable steady state to more complicated geometries. With decreasing stiffness, the first non-flat conformation appeared is a C shape, or a ``taco" shape (Fig.~\ref{fig:resc_flex_evo}(a)) as the circular sheet folds onto itself. The next conformation observed is an S shape, as shown in Fig.~\ref{fig:resc_flex_evo}(b). This shape and some of the other wrinkled shapes slowly rotate around the $z$ axis during flow, due to the asymmetry of the conformation. For even smaller $\kb$, the sheet develops into more complicated structures, such as ``heart" and ``double S", as shown in Fig.~\ref{fig:resc_flex_evo}(c) and Fig.~\ref{fig:resc_flex_evo}(d), respectively. A full parameter sweep of results for a range of $\Ca$ and $\kb$ will be given below, after we illustrate the various types of conformations that we observe.

\begin{figure}[h]
    \centering
    \captionsetup{justification=raggedright}
    \includegraphics[width=\textwidth]{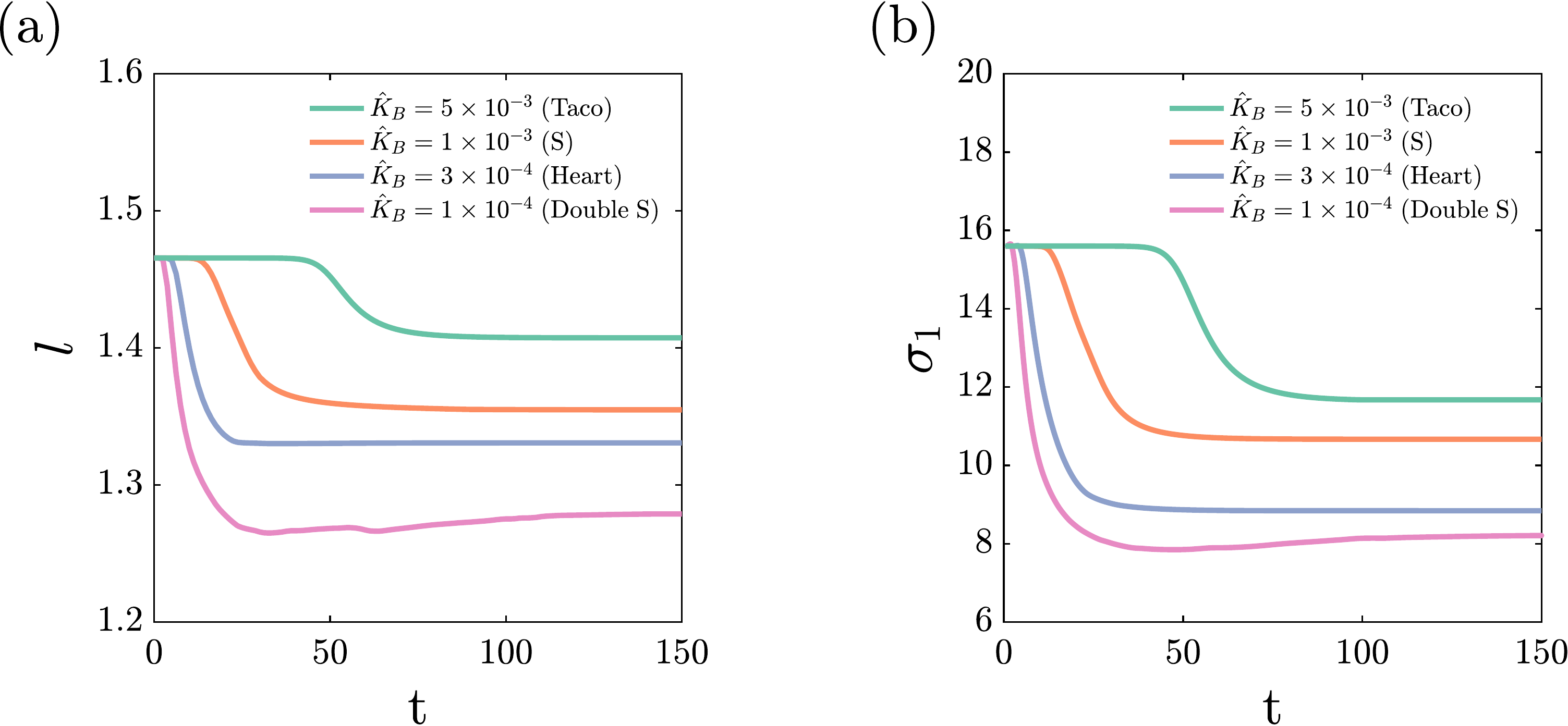}
    \caption{(a) Transient stretched length evolution of different conformations in Figure \ref{fig:resc_flex_evo} with fixed $\Ca = 0.25$. (b) Largest eigenvalue $\sigma_{1}$ evolution of the cases in Figure \ref{fig:resc_stresslet}(a).}
    \label{fig:resc_stresslet}
\end{figure}

Since the sheets with different $\kb$ develop into various conformations, we here illustrate how different conformations affect the stretching dynamics from two perspectives: the stretched length of the sheet and the force dipole exerted by the sheet on the flow. 
Figure \ref{fig:resc_stresslet}(a) shows the transient stretched length $l$ for the cases presented in Fig.~\ref{fig:resc_flex_evo}. All sheets begin with the same flat steady state (fixed $\Ca$) and a small random perturbation. Interestingly, we observe that as the sheet begins to wrinkle, the induced out-of-plane deformation leads to a decrease in stretched size. The decrease in $l$ can be easily observed in Fig.~\ref{fig:resc_flex_evo}. From Fig.~\ref{fig:resc_stresslet}(a), sheets with smaller $\kb$ display a more significant decrease in stretched length, and the decrease in the stretched length is associated with the type of conformation observed. For instance, as the taco folds onto itself, it ``embraces" and traps fluid inside the taco, screening it from the fluid motion outside of the taco. 

\begin{figure}[t!]
    \centering
    \captionsetup{justification=raggedright}
    \includegraphics[width=\textwidth]{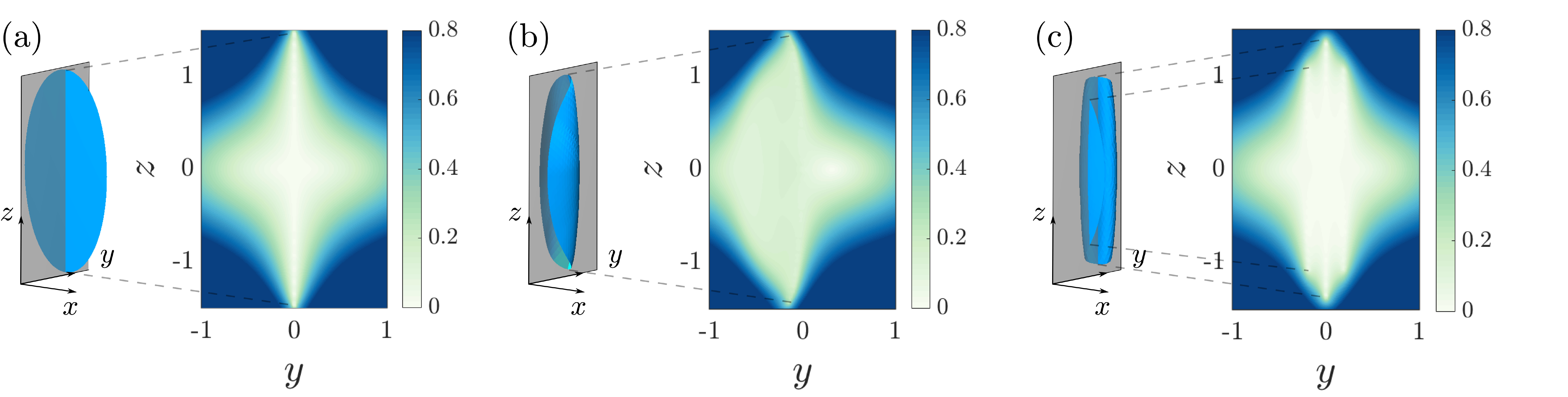}
    \caption{Velocity norm $||\mathbf{v}||$ on a slice (plane $x$ = 0) through the steady states at $\Ca = 0.25$: (a) A flat sheet as at $\mathrm{t} = 0$ in Fig.~\ref{fig:resc_flex_evo}. (b) The taco shape in Fig.~\ref{fig:resc_flex_evo}(a). (c) The S shape in Fig.~\ref{fig:resc_flex_evo}(b).}
    \label{fig:resc_hi}
\end{figure}

To visualize the effect of this hydrodynamic screening of the interior of the taco from the imposed flow, we examine the velocity on a symmetry plane cutting vertically through the taco shape. Figure \ref{fig:resc_hi}b shows this plane and the velocity magnitude $||\mathbf{v}||$ on it. We show a slice through the flat steady state in Fig.~\ref{fig:resc_hi}a as a reference, where the velocity field is similar to the bulk extensional flow. By comparing with the flat case, we observe that the velocity inside the taco is smaller than outside, indicating hydrodynamic screening of the interior. In Fig~\ref{fig:resc_hi}c, we also present results for an S conformation. Here the decrease in velocity of the fluid inside the wrinkles is even more apparent than in the taco case. 
Because of this effect, the sheet is less exposed to the extensional flow, leading to the decrease in length. This hydrodynamic screening applies as well for the other conformations: for the heart or double S shapes, fluid is trapped in the wrinkles. Those conformations have different stretched lengths because of the different extents of hydrodynamic screening. 

The deformation of the sheets is closely related to the stress that they exert on the fluid, so it is also of interest to examine the force dipole associated with the deforming sheets. (The total stress associated with the sheets is simply the force dipole per sheet times the number density in suspension.) The force dipole is given by:
\begin{equation}
    \mathbf{D} = -\sum_i\mathbf{F}_i\mathbf{x}_i,
\end{equation}
where $\mathbf{F}_i$ is the force exerted on the fluid by node $i$ on the sheet. The eigenvalues $\sigma_{i}$ of $\mathbf{D}$ reveal the strength of the dipole along the three principal directions. Here, the first eigenvalue $\sigma_{1}$ indicates the dipole acting along the $z$ axis (flow direction). The second and third eigenvalues correspond to two orthogonal dipoles on the $x-y$ plane.
In Fig.~\ref{fig:resc_stresslet}(b), we plot evolution of $\sigma_{1}$ for same cases from conformations from Fig.~\ref{fig:resc_stresslet}(a). Here, we focus on $\sigma_{1}$,  as the two in-plane eigenvalues are negative and small in magnitude.
The evolution of $\sigma_{1}$ follows that of stretched length. The magnitude of $\sigma_{1}$ relates to the stretched length of the sheet because more stretched sheets exert stronger force dipole along the flow direction. 

\begin{figure}[h]
    \centering
    \captionsetup{justification=raggedright}
    \includegraphics[width=1\textwidth]{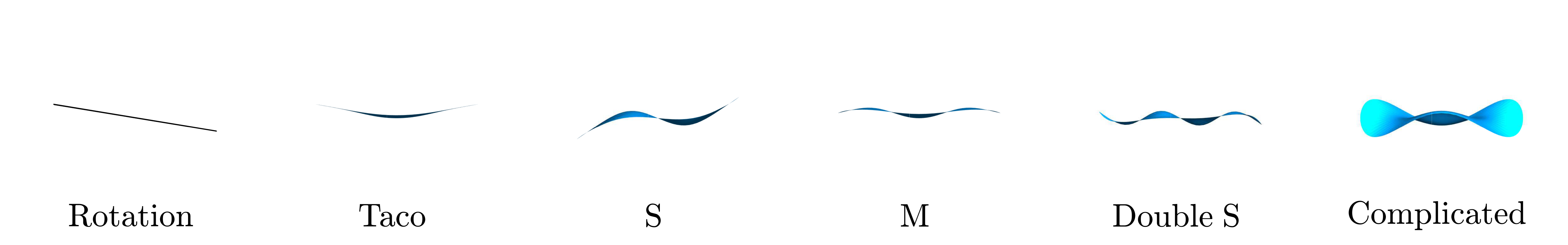} 
    \caption{Top view of examples of deformation modes (added to the flat steady state) obtained from linear stability analysis.}
    \label{fig:resc_modes}
\end{figure}

To further understand the relation between conformation and deformability, we perform a linear stability analysis to examine the response of the sheet to infinitesimal perturbations.
We describe the conformation of a sheet by stacking all the nodal positions into a vector $\mathbf{X}$. The steady state, a flat sheet, is described by $\mathbf{X}_s$. The evolution of the sheet is described by 
\begin{equation}
    \dot{\mathbf{X}} = f(\mathbf{X}),
    \label{eq:stability1}
\end{equation}
where $f(\mathbf{X})$ is given by our numerical simulation algorithm described above. 
We first apply an infinitesimal perturbation $\varepsilon\hat{\mathbf{X}}$ along the out-of plane direction of the steady state $\mathbf{X}_s$, where $\varepsilon$ is the magnitude of the perturbation and $\hat{\mathbf{X}}$ represents a deformation pattern:
\begin{equation}
     \mathbf{X} = \mathbf{X}_s + \varepsilon\hat{\mathbf{X}}.
\end{equation}
If $\varepsilon \ll 1$, we assume that dynamics remain in the linear regime. We can Taylor-expand \eqref{eq:stability1} around $\mathbf{X}_s$, note that $\dot{\mathbf{X}}_s=f(\mathbf{X}_s)=0$, and ignore higher order terms in $\varepsilon$ to find that
\begin{equation}
    \dot{\hat{\mathbf{X}}} = \mathbf{L}\hat{\mathbf{X}}.
\end{equation}
This equation has solution:
\begin{equation}
    \hat{\mathbf{X}}_{t+\tau} = e^{\mathbf{L}\tau}\hat{\mathbf{X}}_{t} = \mathbf{A}\hat{\mathbf{X}}_{t}.
    \label{eq:stability2}
\end{equation}
We determine $\mathbf{A}\hat{\mathbf{X}}_{t}$ by performing simulations with our full nonlinear simulation method in the case where $\varepsilon$ is very small and the dynamics are approximately linear. The Arnoldi method \cite{trefethen1997} is then used to determine the dominant eigenvalues and eigenvectors (modes) of $\mathbf{A}$. {In the analysis, we take $\tau = 0.05$ strain units and we need less than $200$ iterations to get convergence of the dominant eigenvalues.}
The matrices $\mathbf{A}$ and $\mathbf{L}$ have same the eigenvectors, and eigenvalues $\lambda_L$ of $\mathbf{L}$ are related to those of $\mathbf{A}$ ($\lambda_A$) by $\lambda_A = e^{\lambda_L\tau}$. If a mode has an eigenvalue with a positive real part ($\mathrm{Re}(\lambda_L)>0$), the mode is unstable and its amplitude as a component of the initial condition will grow exponentially until nonlinear effects become important. A mode with eigenvalue that has a negative real part ($\mathrm{Re}(\lambda_L)<0$) will decay exponentially. Figure \ref{fig:resc_modes} shows some typical modes arising in our analysis and the names we have given them based on their shapes. One important mode (leftmost in Fig.~\ref{fig:resc_modes}) is the rotational invariance mode, which always has a zero eigenvalue ($\lambda_L=0$). It arises because the shape dynamics are independent of orientation around the $z$ axis.

\begin{figure}[h]
    \centering
    \captionsetup{justification=raggedright}
    \includegraphics[width=0.85\textwidth]{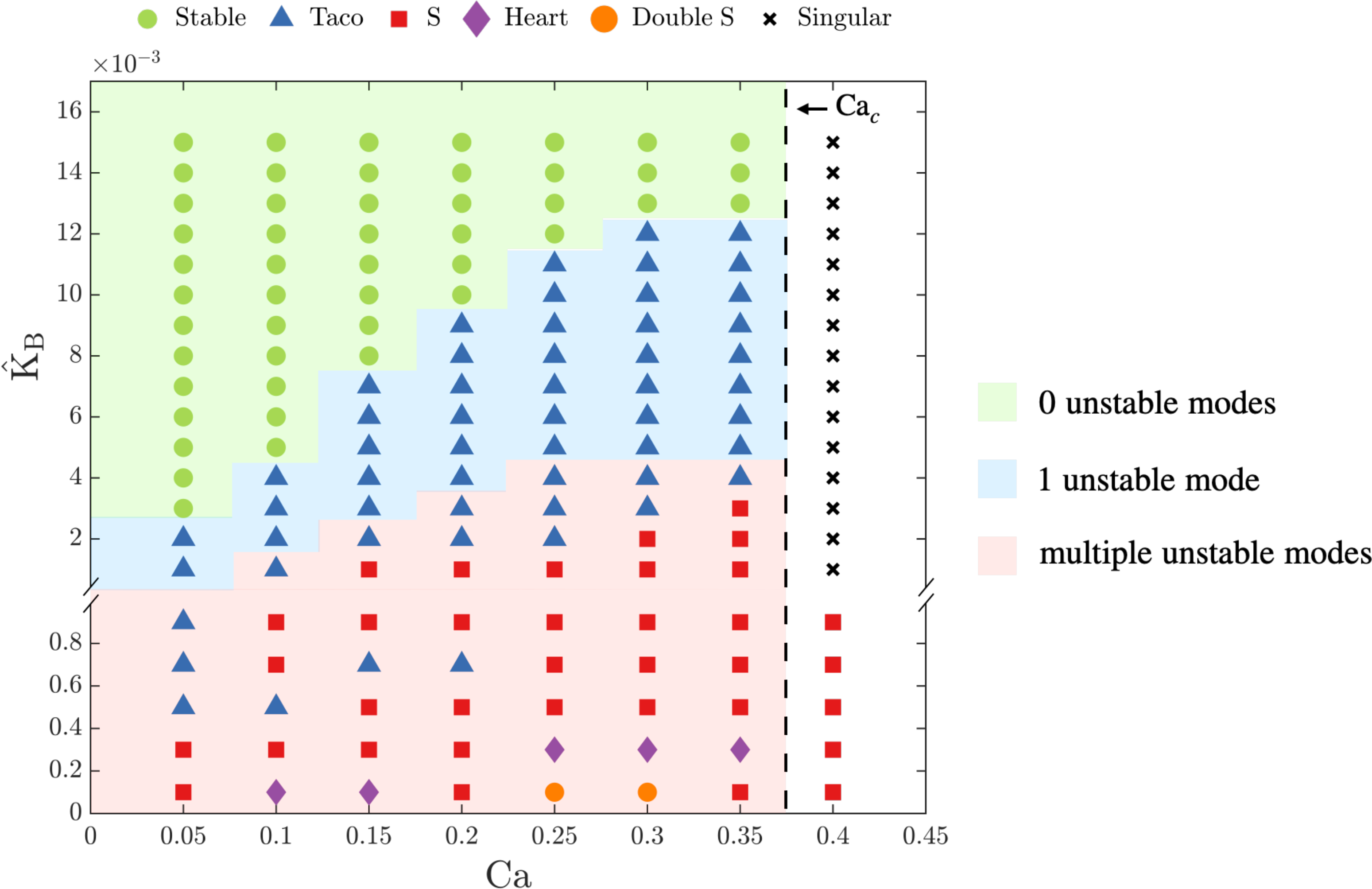}
    \caption{Simulation results and linear stability analysis for a disc with neo-Hookean model ($c = 0$). The symbols represent the final conformation from numerical simulation, where the initial condition is a compact and flat sheet under random perturbation, except for the sheets above $\Ca_c$. The color of the shaded region indicates the number of unstable modes obtained from linear stability analysis.}
    \label{fig:res_compact_phase_diagram}    
\end{figure}

We now apply the linear stability analysis to help understand the numerical simulation results by performing a parameter sweep of $\Ca$ and $\kb$. Figure \ref{fig:res_compact_phase_diagram} consists of two parts: the shaded area with different colors are results from linear stability analysis, showing numbers of unstable modes found for sheets with different $\Ca$ and $\kb$; the markers represent final conformations from numerical simulation. 
For nonlinear simulations, the initial condition chosen is a flat (compact) steady state at the corresponding $\Ca$ (the steady state obtained from a stiff sheet) subjected to a small random perturbation. If no compact steady state is available (the special cases above $\Ca_c$), we start from a sheet at equilibrium with randomly perturbed surface.

The results in Fig.~\ref{fig:res_compact_phase_diagram} are divided into three parts based on the number of unstable modes found. The green area shows the regime where no unstable modes are found, so linear stability analysis predicts that the sheet relaxes to the flat steady state under infinitesimal perturbation. 
For those sheets with no unstable modes, their final conformation with finite random initial conditions from numerical simulation also stays flat. We only introduce perturbations at $t=0$, so if a sheet evolves to the flat state it will stay there forever.

In the blue regime, the stability analysis gives one unstable mode, the taco mode. 
The numerical simulation agrees with what we understand from linear stability analysis: if the taco is the only unstable mode, it is the only unstable deformation that the sheet evolves into, as the sheet is stable under all other types of perturbation. Therefore, the taco is the only stable conformation in the long-term. The linear stability analysis also can predict the final conformation in this regime when only one unstable mode emerges.

\begin{figure}[t!]
    \centering
    \captionsetup{justification=raggedright}
    \includegraphics[width=\textwidth]{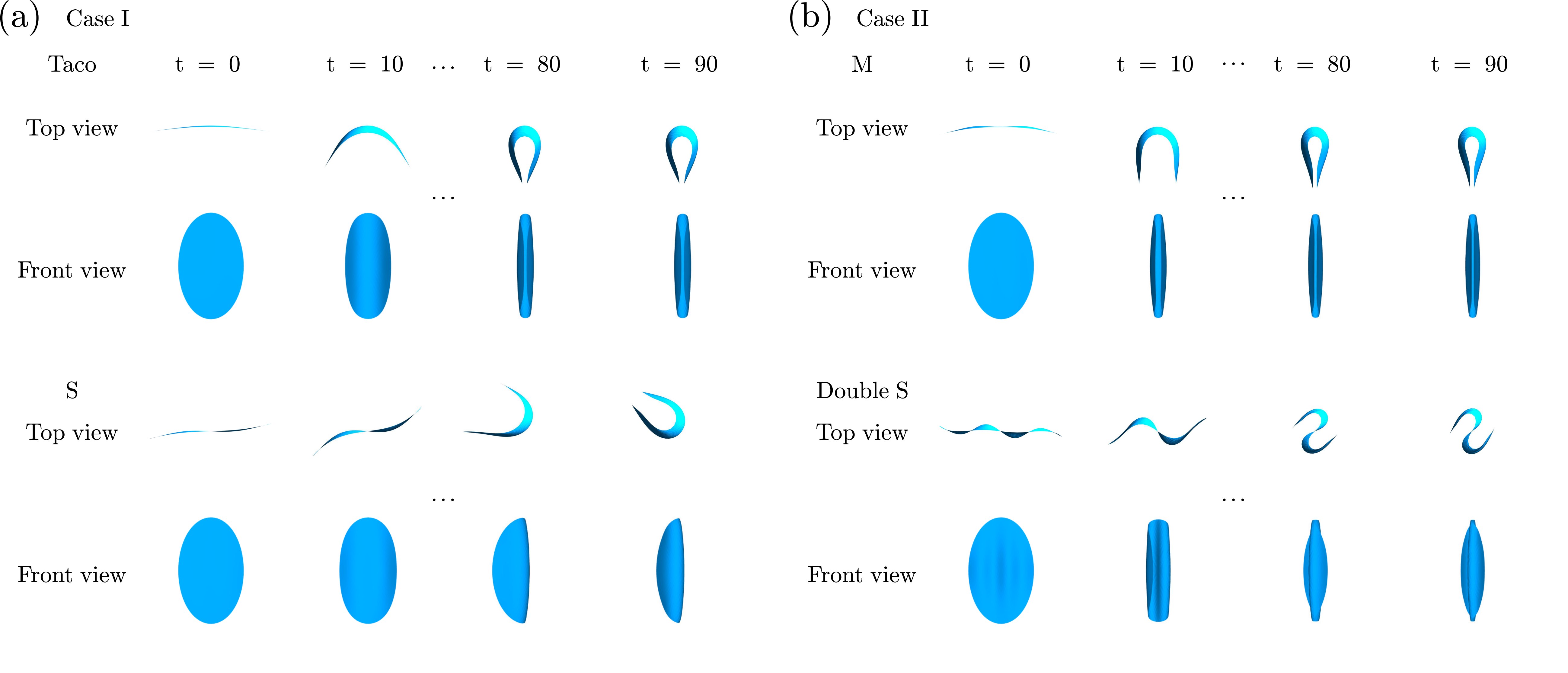}
    \caption{Top view and front view of evolutions of modes when they become unstable, with their short-term and long-term dynamics. (a) Taco and S with $\Ca = 0.2$, $\kb = 3\times 10^{-3}$. (b) M and double S with $\Ca = 0.2$, $\kb = 1\times 10^{-3}$. See Supplemental Material at [URL will be inserted by publisher: Appendix \ref{sec:appendix} Movie 2 and 3] for animated movies.}
    \label{fig:resc_modes_evolution}
\end{figure}

When it comes to the red area, the presence of multiple unstable modes implies that various potential deformations can be obtained, depending on the initial condition. Here, the numerical simulation also suggests many complicated conformations. From the phase diagram, the simulation shows a transition from taco to more complicated S, heart and double S with decreasing $\kb$. In order to understand these complicated scenarios, we present two examples in this regime to help illustrate the relation between unstable modes and final conformation. Figure \ref{fig:resc_modes_evolution} shows the shape evolution of the unstable modes fed as the initial condition in two cases. The first case ($\Ca = 0.2$, $\kb = 3 \times 10^{-3}$) only has two unstable modes, and is located at the border of red and blue area. The second case ($\Ca = 0.2$, $\kb = 1 \times 10^{-3}$) has more than two unstable modes, and we only show two of them.

Based on the evolution in Fig.~\ref{fig:resc_modes_evolution}(a), the first case, the long time dynamics from either taco or S initial conditions evolve into a taco. This observation shows the limit of the linear analysis that it is unable to precisely determine the long term dynamics of a nonlinear simulation. 
The second case, Fig.~\ref{fig:resc_modes_evolution}(b), is more complex, with multiple unstable modes that show different unstable wrinkling patterns. The mirror symmetrical mode M and rotationally symmetric mode double S keep their symmetry for both the short-term and long-term dynamics. Therefore, the mode evolution also implies that the long-time sheet conformations in the red area may vary based on the initial condition, especially for cases with small $\kb$. 

\begin{figure}[h]
    \centering
    \captionsetup{justification=raggedright}
    \includegraphics[width=\textwidth]{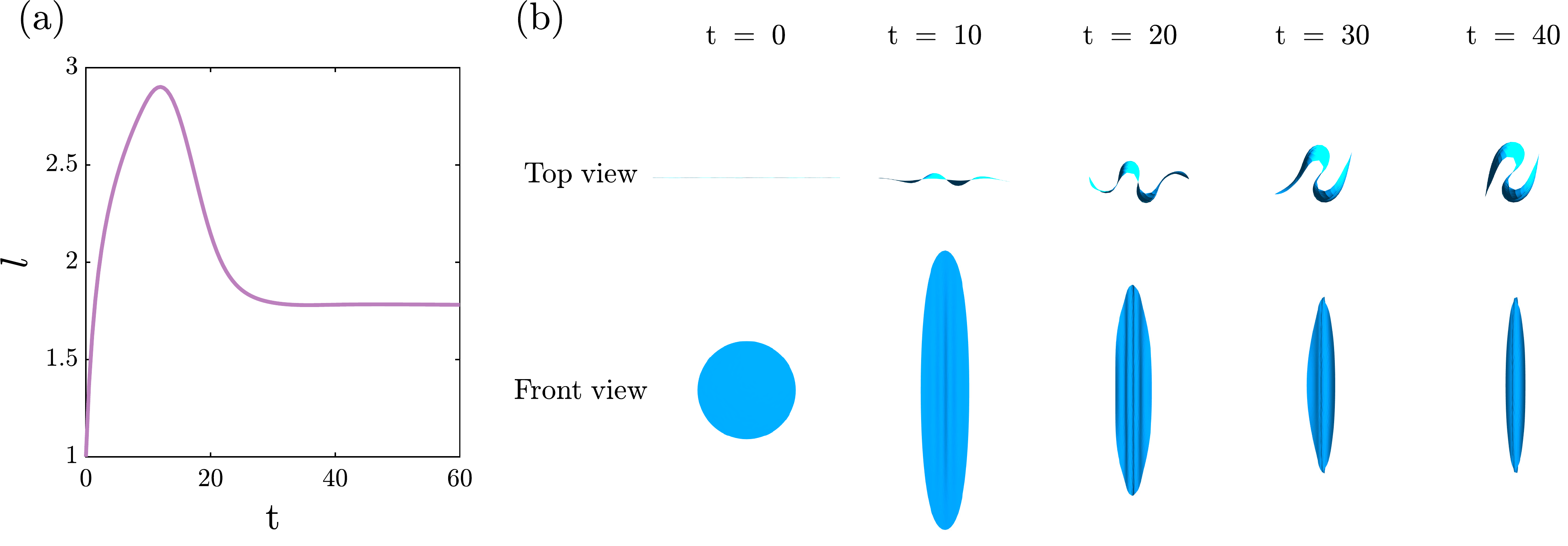}
    \caption{Dynamics of a flexible sheet for $\Ca > \Ca_c$ for the neo-Hookean model ($c = 0$): $\Ca = 0.4$, $\kb = 5 \times 10^{-4}$. (a) Evolution of length. (b) Snapshots of evolution. See Supplemental Material at [URL will be inserted by publisher: Appendix \ref{sec:appendix} Movie 4] for animated movies.}
    \label{fig:resc_singular_evo}
\end{figure}

We observed above that the wrinkling instability leads to a decrease in the steady-state length. We conclude this section with an interesting extreme example of this phenomenon. If $\Ca$ exceeds $\Ca_c$, there is no steady state at all for the flat neo-Hookean sheet: the sheet stretches indefinitely. Nevertheless, if the bending stiffness is sufficiently small that the sheet wrinkles, the wrinkled sheet will reach a finite-length steady state. This observation is indicated previously in Fig.~\ref{fig:res_compact_phase_diagram} for $\Ca = 0.4$. Figure \ref{fig:resc_singular_evo}(a) shows the length evolution for a case where this happens, and Fig.~\ref{fig:resc_singular_evo}(b) shows the corresponding shape evolution.  
The sheet is at rest at $t=0$ when flow begins. Initially the length increases, but once the sheet begins to wrinkle, $l$ decreases then reaches a steady state value.
The wrinkling leads to hydrodynamic screening of the sheet surface in the folds, reducing the local drag on the sheet and causing it to reach a finite steady state value of $l$. 
These results imply that the actual $\Ca_c$ might vary with $\kb$ in the compact-stretched transition, as we discuss in the next section.

    \subsection{Dynamics of flexible sheets in the stretched (high $\Ca$) regime and the compact-stretched transition} \label{sec:results_stretched}
      

In this section, we first continue the discussion of wrinkling with a focus on the stretched branch in Fig.~\ref{fig:resu_stiff_csh}(b). Then, we address the influence of wrinkling on the compact-stretched transition and bistability.
 
\begin{figure}[h]
    \centering
    \captionsetup{justification=raggedright}
    \includegraphics[width=0.8\textwidth]{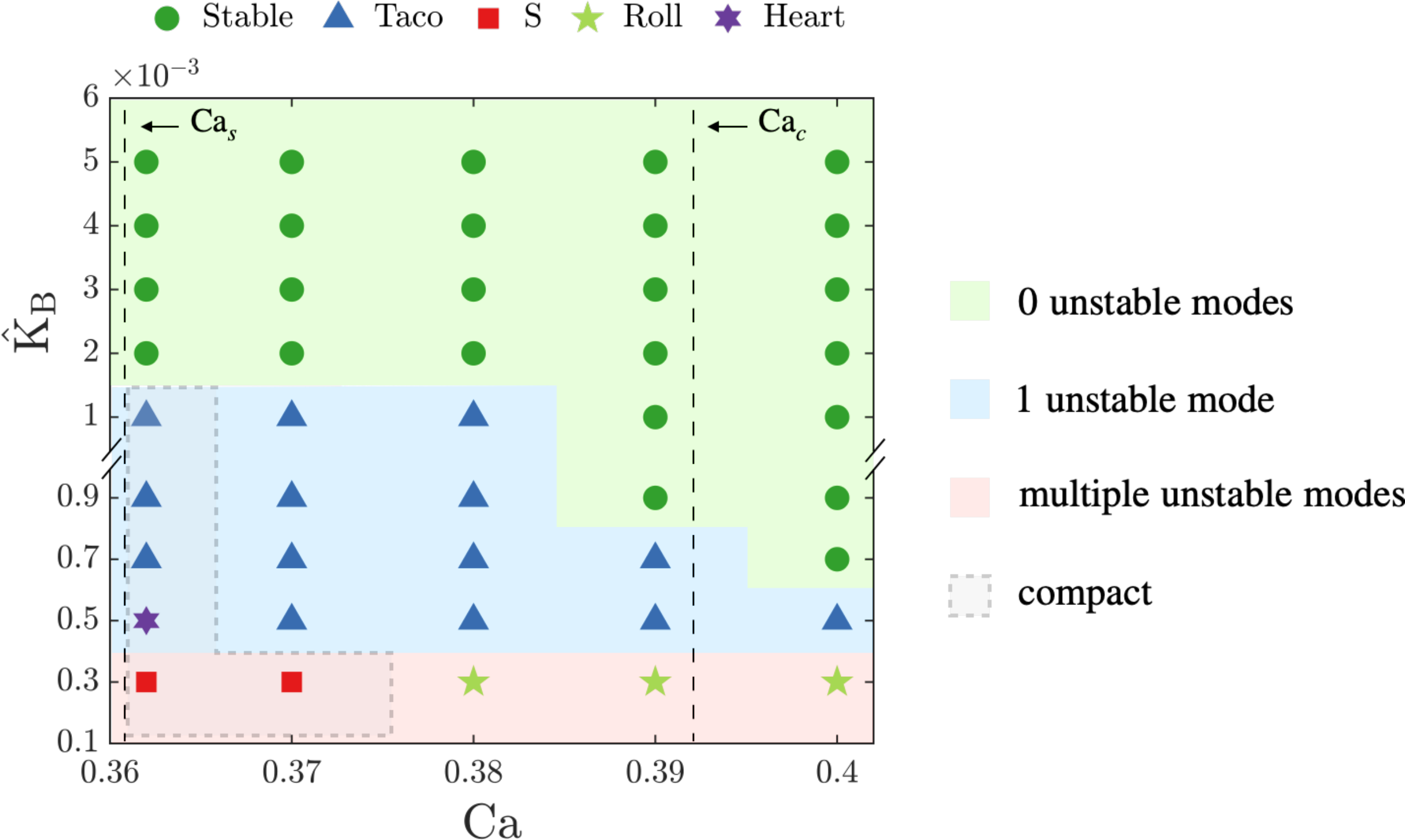}
    \caption{Simulation results and linear stability analysis for a disc  for parameters encompassing the bistable regime for the flat sheet ($c = 1\times 10^{-5}$). The symbols represent the final conformation from numerical simulation, where the initial condition is a stretched and flat sheet under random perturbation. The boxed region indicated where an initially stretch state reverts to a compact state upon wrinkling. The color of the shaded region indicates the number of unstable modes obtained from linear stability analysis.}
    \label{fig:res_stretched_phase_diagram}    
\end{figure}
 
When a sheet is highly stretched, it experiences strong shear stresses due to its increased length and surface area, which further affects the conformation. We performed a parameter sweep focusing on the stretched $\Ca$ regime in Fig.~\ref{fig:resu_stiff_csh}(b) and combined the linear stability analysis to examine the influence of $\kb$, and thus the wrinkling instability, on the conformation. Figure \ref{fig:res_stretched_phase_diagram} illustrates the final conformation of the sheet with different $\kb$, as well as the number of unstable modes obtained from linear stability analysis. 
Compared with Fig.~\ref{fig:res_compact_phase_diagram} for compact sheets, Fig.~\ref{fig:res_stretched_phase_diagram} indicates that stretched sheets are more resistant than compact sheets at same parameter values to out-of-plane deformation when $\kb$ is relatively large: a compact sheet wrinkles while a stretched sheet with same parameters stays flat. Interestingly, in contrast to the compact regime, where sheets with larger $\Ca$ deform first, stretched sheets with smaller $\Ca$ first evolve into non-flat conformations while stretched sheets with larger $\Ca$ stay flat. 

We observed that upon decreasing $\kb$, the stretched sheets close to $\Ca_s$ may snap to the compact branch after wrinkling, exhibiting a significant decrease in stretched length. Due to hydrodynamic screening (reduced drag) arising from wrinkling, a fully stretched shape can no longer be maintained. We indicate those cases in Fig.~\ref{fig:res_stretched_phase_diagram} as a boxed area. This observation implies an increase in $\Ca_s$ for smaller $\kb$. 
 
\begin{figure}[t!]
    \centering
    \captionsetup{justification=raggedright}
    \includegraphics[width=\textwidth]{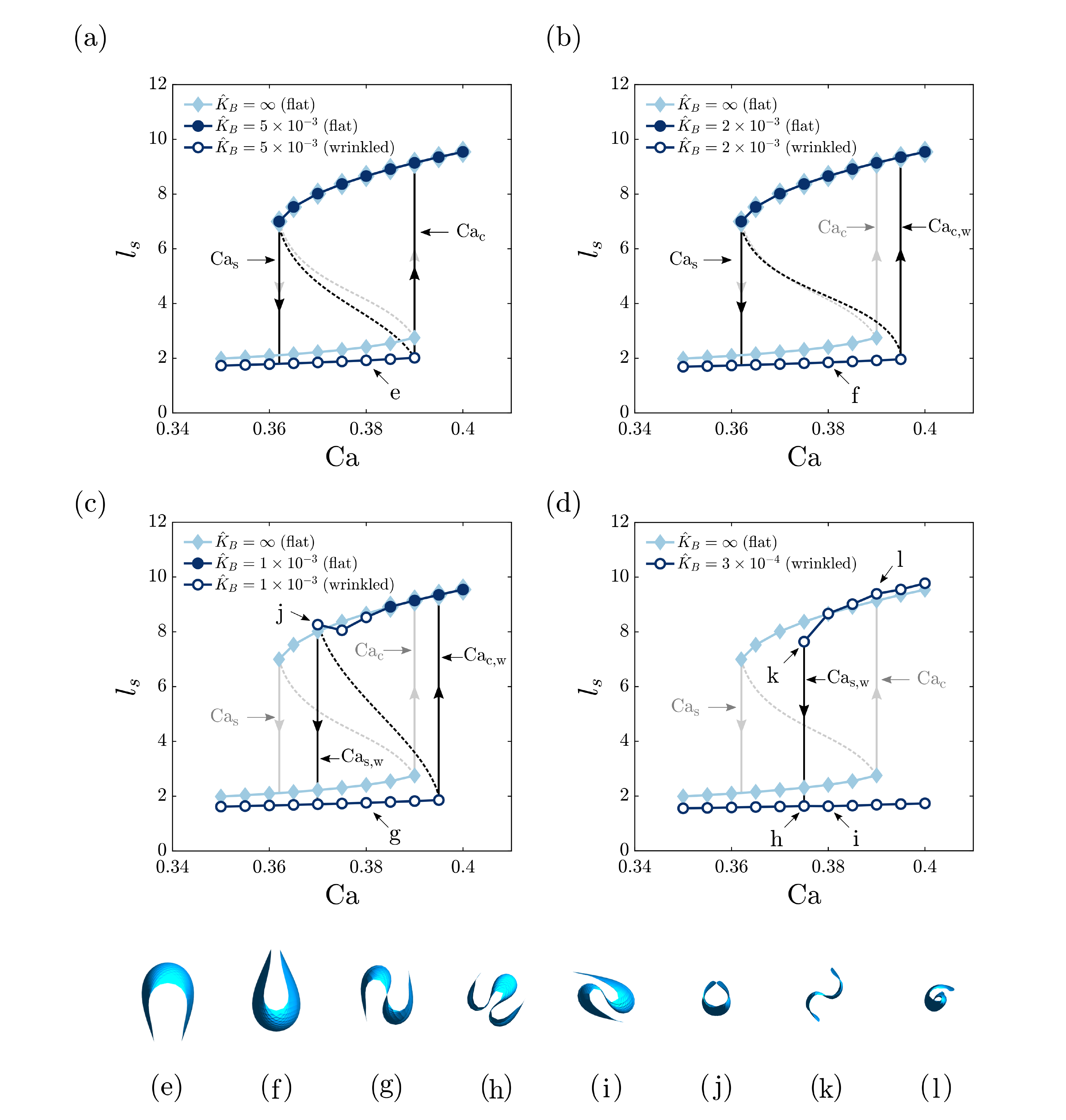}
    \caption{(a)-(d): Bifurcation diagram for a flexible disc ($c = 1\times 10^{-5}$). For reference, the light blue data indicate results for a flat stiff sheet. The hollow data points show a wrinkled sheet and the filled points represent a flat state. (a) $\kb = 5\times 10^{-3}$. (b) $\kb = 2\times 10^{-3}$. (c) $\kb = 1\times 10^{-3}$. (d) $\kb = 3\times 10^{-4}$. Labels $\Ca_{c,w}$ or $\Ca_{s,w}$ denote critical capillary numbers for the wrinkled case. (e)-(l): Conformations at the points labeled e - l on the bifurcation diagrams.}
    \label{fig:ress_flex_csh}    
\end{figure}

We also performed linear stability analysis on the stretched steady states and split Fig.~\ref{fig:res_stretched_phase_diagram} into three parts based on the number of unstable modes. The stability analysis results agree with simulation in identifying the regime where sheets stay flat and where sheets become taco-shaped. For sheets with multiple unstable modes, the final conformation again depends on the initial conditions -- multiple stable final wrinkled states are possible. 
{Figure \ref{fig:res_stretched_phase_diagram} shows that initially flat sheets with small $\kb$ can form asymmetric shapes like the roll that is shown in Fig.~\ref{fig:ress_flex_csh}. These shapes slowly rotate due to their asymmetry, and the final conformations deviate significantly from the shape of the unstable modes from linear analysis.}



To summarize the influence of wrinkling on the bistability, we present in Fig.~\ref{fig:ress_flex_csh} the bifurcation diagrams of stretch vs. $\Ca$ for four different values of $\kb$, showing results for flat sheets and sheets allowed to wrinkle. Figure \ref{fig:ress_flex_csh}(a) shows the bifurcation diagram for a relatively stiff sheet ($\kb = 5\times 10^{-3}$): all cases on the stretched branch stay flat, while the compact branch wrinkles, displaying a taco shape. Figure \ref{fig:ress_flex_csh}(e) presents a snapshot of the taco shape that is found at the point labeled `e' on Fig.~\ref{fig:ress_flex_csh}(a). At this stage, both $\Ca_c$ and $\Ca_s$ are close to the values found for the flat case.

If we decrease $\kb$ to $2\times 10^{-3}$ (Fig.~\ref{fig:ress_flex_csh}(b)), the $\Ca_c$ for the compact branch begins to shift to higher $\Ca$ due to wrinkling, and we denote the new critical capillary number $\Ca_{c,w}$. The interval between $\Ca_c$ and $\Ca_{c,w}$ is similar to the case we described above for the neo-Hookean model, in which cases above the singularity can maintain a compact state due to wrinkling-induced hydrodynamic screening. {For this parameter set, the steady shapes on the compact branch are tacos (Fig.~\ref{fig:ress_flex_csh}(f))} and the stretched branch remains flat. 

When $\kb = 1\times 10^{-3}$ (Fig.~\ref{fig:ress_flex_csh}(c)), we observe that some stretched sheets also wrinkle, {which induces a transition back to a compact state}, resulting in a shift of the upper saddle-node bifurcation to a larger value of $\Ca$ that we denote $\Ca_{s,w}$. Here the conformations on that compact branch all become a rotating S, as shown in Fig.~\ref{fig:ress_flex_csh}(g). For the stretched branch, the deformed sheets form a taco (Fig.~\ref{fig:ress_flex_csh}(j)).
 
Finally, Fig.~\ref{fig:ress_flex_csh}(d) shows a case, $\kb = 3\times 10^{-4}$, where both the upper and lower critical capillary numbers strongly deviate from the stiff case. Here, we did not determine the new $\Ca_c$ specifically as it is above the parameter range we have considered. The compact branch exhibits two different conformations: heart (Fig.~\ref{fig:ress_flex_csh}(h)) for $\Ca< 0.38$ and S (Fig.~\ref{fig:ress_flex_csh}(i)) otherwise. On the stretched branch, all cases evolve into two non-flat conformations. Except the case $\Ca = 0.375$, which becomes a S (Fig.~\ref{fig:ress_flex_csh}(k)), all other cases become a slowly rotating roll shape (Fig.~\ref{fig:ress_flex_csh}(l)). In summary, based on what we have observed, decreasing $\kb$ strongly modifies the compact-stretched transition by shifting saddle-node bifurcations to higher $\Ca$.


\section{CONCLUSION} \label{sec:conclusion}
  We have applied numerical simulation and linear stability analysis to systematically explore the compact-stretched transition of soft elastic sheets in uniaxial extensional flow.

For stiff sheets ($\kb = \infty$) modeled with the neo-Hookean model, we observe a singularity in conformation at a critical capillary number $\Ca_c$. Sheets modeled with the Yeoh model and small values of the nonlinearity paramter $c$ exhibit a discontinuous transition to a highly stretched state beyond $\Ca_c$ --  a \emph{compact-stretched} transition. The discontinuity in size marks a bistable regime (defined by $\Ca_s$ and $\Ca_c$), where both compact and stretched states can exist based on deformation history.

For flexible sheets, wrinkling instability strongly modifies the compact-stretched transition and bistability. 
In the compact regime, we observe various wrinkled conformations with different sizes, due to different extents of hydrodynamic screening: the sheet traps the fluid inside wrinkles, leading to a decrease in stretched length. With sufficiently small $\kb$, cases above $\Ca_c$ can stay in  a compact conformation without becoming fully stretched, resulting in a larger $\Ca_c$ for wrinkled sheets. We develop a simple linear stability analysis that can predict some of the conformations.

In the stretched regime, the linear stability analysis can also predict final conformation of stretched sheets with limited unstable modes. Compared with the compact sheets at same parameter values, the stretched sheets are more resistant to out-of plane deformation. Highly stretched flat sheets with small $\kb$ may become wrinkle and snap to compact conformations, yielding a larger $\Ca_s$. Therefore, we find the decrease in $\kb$ shifts the bistable regime to higher $\Ca$.

This study deepens our fundamental understanding of the dynamics of a soft sheet-like particles in uniaxial extensional flow and displays important interactions between the compact-stretched transition and wrinkling instabilities. The broad parameter range considered here covers sheets with different in-plane deformability and out-of-plane flexibility, ranging from flexible nanosheets to extensible polymer films. We believe the current work will help guide design of processes for flow-controlled assembly and synthesis of sheet-like particles. 


\begin{acknowledgments}
  
This material is based on work supported by the National Science Foundation under grant No.~CBET-1604767. We acknowledge Sarit Dutta for useful comments.
\end{acknowledgments}

\begin{appendix}
\appendix 
\section{Movies} \label{sec:appendix}
\begin{itemize}
	\item Movie 1: Compact sheets ($c = 0,\; \Ca = 0.25$) with different $\kb$ wrinkle to form different conformations (taco, S, heart, double S, respectively).
	\item Movie 2: Mode evolutions of a compact sheet ($c = 0,\; \Ca = 0.2,\; \kb = 3 \times 10^{-3}$).	
	\item Movie 3: Mode evolutions of a compact sheet ($c = 0,\; \Ca = 0.2,\; \kb = 1 \times 10^{-3}$).
	\item Movie 4: A compact sheet above singularity ($c = 0,\; \Ca = 0.4,\; \kb = 5 \times 10^{-4}$) wrinkles to form an S shape.
\end{itemize}

\end{appendix}

\section*{REFERENCES}
\bibliography{uniaxial_cite}

\end{document}